\documentclass[epj,twocolumn]{webofc}
\usepackage[varg]{txfonts}   
\usepackage{color}
\newcommand{\nc}{\newcommand}
\nc{\scm}{\scriptscriptstyle\mathrm}
\nc{\f}{\frac}
\nc{\eps}{\varepsilon}
\nc{\vp}{\varphi}
\nc{\tvp}{\widetilde{\varphi}}
\nc{\D}{\mbox{$\not\!\!D$}}
\nc{\Db}{\mbox{${\raisebox{2mm}{\boldmath ${}^\leftarrow$}\hspace{-4mm} D}$}}
\nc{\Dfb}{\mbox{$\raisebox{2mm}{\boldmath ${}^\leftrightarrow$}\hspace{-4mm} D$}}
\nc{\vpj }{\mbox{${\vp^\dag i\, \raisebox{1.5mm}{\boldmath ${}^\leftrightarrow$}\hspace{-3mm} D_\mu\,\vp}$}}
\nc{\vpjt}{\mbox{${\vp^\dag i\, \raisebox{1.5mm}{\boldmath ${}^\leftrightarrow$}\hspace{-3mm} D_\mu^{\,I}\,\vp}$}}
\nc{\btb}{\begin{tabular}}    \nc{\etb}{\end{tabular}}
\newcommand{\Eqn}[1]{Eq.~(\ref{#1})}

\woctitle{Flavour changing and conserving processes}
\begin{document}
\title{Lepton-flavour violating decays in theories with dimension 6 operators}
\author{Giovanni Marco Pruna\inst{1}\fnsep\thanks{\email{Giovanni-Marco.Pruna@psi.ch}} \and
        Adrian Signer\inst{1,2}\fnsep\thanks{\email{Adrian.Signer@psi.ch}}
}
\institute{Paul Scherrer Institut, CH-5232 Villigen PSI, Switzerland
\and
           Physik-Institut, Universit\"at Z\"urich, CH-8057 Z\"urich, Switzerland
}
\abstract{Despite a large experimental effort, so far no evidence for
  flavour-violating decays of charged leptons such as $l_i\to
  l_j\gamma$ and $l_i\to l_j l_k l_k$ has been found. The absence of a
  signal puts very severe constraints on many extensions of the
  Standard Model. Here we apply a model independent approach by
  studying such decays in the Standard Model effective field theory.
  Going beyond leading order in the Standard Model couplings and
  considering all dimension 6 operators that might lead to
  lepton-flavour violation, we are able to extract limits on a large
  number of Wilson coefficients of such operators. We are also able to
  compare the impact of particular searches and find, for example,
  that flavour-violating decays of the $Z$-boson $Z\to \mu e$ are much
  more constrained from low-energy experiments $\mu\to e \gamma$ than
  from the limits of current and future direct searches at high
  energy.  }
\maketitle
\section{Introduction}
\label{sec-1}
It is well established that the Standard Model (SM) of particle
physics provides a structure of gauge symmetries that accidentally
conserves the leptonic flavour. This opens up a possibility for very
stringent tests of the SM and powerful searches for physics beyond the
SM. For this reason, a longstanding experimental effort has been
devoted to the search of lepton-flavour violating (LFV) signals both
in the neutral and charged sectors.

In the neutrino sector, the evidence for flavour violation is by now
established beyond any doubt~\cite{Fukuda:1998mi, Ahmad:2001an,
  Ahmad:2002jz}. Thus, the SM with only left-handed neutrinos needs to
be extended. Concerning the charged lepton sector, searches for LFV in
$\mu$ and $\tau$ decays at low energy experiments as well as in $Z$
boson decays at high energy experiments have been analysed for
decades, resulting in ever more stringent limits on various LFV
branching ratios (BRs) of the aforementioned particles. One
possibility to study and interpret the absence (so far) of any charged
LFV signals is to consider specific models beyond the SM and a large
effort has been made in this direction (see e.g.,~\cite{Crivellin:2013wna, Abada:2014kba, Calibbi:2014yha,Gripaios:2014tna, Kersten:2014xaa,
  Beneke:2015lba, Feruglio:2015gka}).

In this proceeding, the adopted strategy follows from a bottom-up
approach established in~\cite{Buchmuller:1985jz} and refined in the
past thirty years~\cite{Grzadkowski:2010es, Lehman:2014jma}. The SM is
considered to be an effective theory valid up to an (unknown) large
energy scale $\Lambda$ and the Lagrangian consists of a systematic
dimensional expansion in $1/\Lambda$ that includes all operators
constructed from SM fields (without right-handed neutrinos) that
respect the SM gauge symmetries, i.e.
\begin{align}\label{eq:bw}
\mathcal{L} =\mathcal{L}_{\mathrm{SM}} + \frac{1}{\Lambda  } \sum_{k} C_k^{(5)} Q_k^{(5)}
+ \frac{1}{\Lambda^2} \sum_{k} C_k^{(6)} Q_k^{(6)} +
\mathcal{O}\left(\frac{1}{\Lambda^3}\right)\, . 
\end{align}
The coefficients $C_k^{(d)}$ are couplings and, very much like the
usual couplings of the SM, they have to be determined by
experiment. Beyond tree level, a renormalisation scheme has to be
chosen to give the couplings a well-defined meaning. 

Once an ultraviolet (UV) complete theory is known that describes the
effects of physics beyond the scale $\Lambda$, the coefficients
$C_k^{(d)}$ of the Lagrangian can be expressed in terms of the
parameters of the UV complete theory. Such a matching has to be done
at the scale $\Lambda$. To obtain the coefficients at the relevant low
energy scale, renormalisation-group (RG) techniques have to be used to
determine the running. We want to stress that due to operator mixing
the RG evolution leads to qualitatively new effects and, at least for
the time being, is not primarily a question of precision.

Performing an expansion in $1/\Lambda$ of the most general
gauge-invariant Lagrangian one is allowed to write only one Dim-$5$
operator~\cite{Petcov:1976ff, Minkowski:1977sc,Weinberg:1979sa}. This
operator provides both neutrino mass terms and LFV in the neutrino
sector. Going beyond tree level, it also causes LFV in the
charged sector. However, the smallness of the neutrino masses results
in very severe constraints on its coefficients. These constraints
in turn imply that the amount of charged LFV provided by the Dim-5
operator is far below any foreseeable experimental capability. Hence,
the next step consists both in the scrutiny of the $19$ Dim-$6$
operators that yield LFV and in the interpretation of the current BR
limits with respect to such extended parameter space.  Given the
complete lack of information on the structure of the UV complete
theory, we make no assumption whatsoever on the relative importance of
the various Dim-$6$ operators.

In the literature, muon decays are often studied using an effective
theory that stops one step before, by integrating out any dynamics at
the electroweak scale. The resulting effective theory is simpler and
contains fewer parameters. However, we refrain from doing this and
work directly with the Lagrangian of Eq.~(\ref{eq:bw}). This provides
us with a framework to combine constraints from low-energy experiments
such as $\mu^+\to e^+\gamma$ and $\mu^+\to e^+e^-e^+$ with LFV
searches at higher energies, such as LFV decays of the $Z$ boson or
$H\to \tau\mu$.

On the low energy side, very accurate limits on the LFV muon decays
$\mu^+\to e^+\gamma$ and $\mu^+\to e^+e^-e^+$ have been made available
by the MEG~\cite{Adam:2013mnn} and SINDRUM~\cite{Bellgardt:1987du}
collaborations, respectively; in the near future, an upgrade of the
MEG experiment~\cite{Baldini:2013ke} will take place almost in
parallel with Mu3e, a new experiment devoted to the $\mu\to 3e$
searches~\cite{Berger:2014vba}. In the tauonic sector, a big effort is
ongoing to continuously improve the bounds on several decay channels
in various experiments~\cite{Aubert:2009ag,Hayasaka:2010np}.

On the high energy side, interesting searches have been performed at
LEP in the context of LFV decays of the $Z$~boson~\cite{Decamp:1991uy,
  Adriani:1993sy, Akers:1995gz, Abreu:1996mj}, and further
improvements have either been made or are expected from the
ATLAS~\cite{Aad:2014bca} and CMS~\cite{CMS:2015hga} experiments at the
LHC collider. Moreover, after the discovery of the Higgs boson~\cite{Aad:2012tfa, Chatrchyan:2012xdj}, new limits on the $H\to
\tau\mu$ BR are currently under investigation~\cite{Aad:2015gha,Khachatryan:2015kon}.

In this proceeding, the constraints on various Dim-6 coefficients are
reviewed in light of the latest bounds on LFV $\mu$ and $\tau$ decays
and in the assumption that the relevant operators are not
correlated. Due to the very good accuracy of the bounds on $l_i\to
l_j\gamma$ transitions, these processes will be considered at the
one-loop level, while the rest of our study will be performed at the
tree level. The tables supplied in this note are neither meant to be
complete nor to exhaust the general discussion on LFV decays in
theories with Dim-6 operators. For example LFV transitions in nuclei,
where a substantial improvement~\cite{Kutschke:2011ux,Kuno:2013mha}
in the experimental sensitivity with respect to the current best
limit~\cite{Bertl:2006up} is expected, is not considered. At this
stage, we also do not include decays $l_i\to l_j\, 2 l_k$ with $k\neq
j$. Furthermore, we also neglect correlations among operators and
potentially important two loop Barr-Zee effects.
The main two messages that will be
delivered in the following sections are: (i) the richness of the
information on the coefficients $C_k^{(d)}$ that can be gained from
low-energy experiments if Eq.~(\ref{eq:bw}) is taken seriously as a
quantum field theory and (ii) the fact that the limits on LFV decays
of the $Z$~boson from low-energy experiments are far more stringent
than those from the high-energy experiments.

Lastly, a remark is required: these proceedings are largely based on the elements collected in Ref.~\cite{Pruna:2014asa}, to which the reader is redirected for further details.
\section{LFV dimension 6 operators}
\label{sec-2}
Among the $19$ Dim-6 operators that trigger LFV transitions, the
relevant ones for the study of $l_i\to l_j\gamma$ (at the one loop
level), $l_i\to 3l_j$,  $Z\to l_il_j$ and $H\to
l_il_j$ are collected in Table~\ref{tab:operators}.

\begin{table}[!ht] 
\centering
\renewcommand{\arraystretch}{1.5}
\begin{tabular}{||c|c||c|c||} 
\hline \hline
\multicolumn{2}{||c||}{$\psi^2 X\vp$} & 
\multicolumn{2}{|c||}{$\psi^2\vp^2 D$}\\
\hline
$Q_{eW}$  & $(\bar l_p \sigma^{\mu\nu} e_r) \tau^I \vp W_{\mu\nu}^I$&  
$Q_{\vp l}^{(1)}$  &$(\vpj)(\bar l_p \gamma^\mu l_r)$\\
$Q_{eB}$  & $(\bar l_p \sigma^{\mu\nu} e_r) \vp B_{\mu\nu}$&   
$Q_{\vp l}^{(3)}$ & $(\vpjt)(\bar l_p \tau^I \gamma^\mu l_r)$\\
 & &    
$Q_{\vp e}$  & $(\vpj)(\bar e_p \gamma^\mu e_r)$ \\
\hline \hline
\multicolumn{4}{||c||}{$\psi^2\vp^3$}\\
\hline
\multicolumn{2}{||c|}{$Q_{e\vp}$} & 
\multicolumn{2}{|c||}{$(\vp^\dag \vp)(\bar l_p e_r \vp)$}\\
\hline\hline
\multicolumn{2}{||c||}{$(\bar ll)(\bar ll)$} & 
\multicolumn{2}{|c||}{$(\bar ll)(\bar qq)$}\\
\hline
$Q_{ll}$    & $(\bar l_p \gamma_\mu l_r)(\bar l_s \gamma^\mu l_t)$ & $Q_{lequ}^{(1)}$&$(\bar l_p^j e_r) \eps_{jk} (\bar q_s^k u_t)$ \\ 
$Q_{ee}$    & $(\bar e_p \gamma_\mu e_r)(\bar e_s \gamma^\mu e_t)$ & $Q_{lequ}^{(3)}$&$(\bar l_p^j \sigma_{\mu\nu} e_r) \eps_{jk} (\bar q_s^k \sigma^{\mu\nu} u_t)$ \\
$Q_{le}$    & $(\bar l_p \gamma_\mu l_r)(\bar e_s \gamma^\mu e_t)$ & \ & \ \\
\hline \hline
\end{tabular}
\caption{LFV Dim-6 operators.\label{tab:operators}}
\end{table}

Working in the physical basis rather than in the gauge basis, the dipole
operators (set $\psi^2 X\vp$) are rewritten using
\begin{align}
Q_{eB}&\rightarrow Q_{e\gamma}c_W-Q_{eZ}s_W,\\
Q_{eW}&\rightarrow -Q_{e\gamma}s_W-Q_{eZ}c_W,
\end{align}
where $s_W=\sin(\theta_W)$ and $c_W=\cos(\theta_W)$ are the sine and
cosine of the weak mixing angle. The term
\begin{align}
\mathcal{L}_{e\gamma} \equiv 
\frac{C_{e\gamma}}{\Lambda^2} Q_{e\gamma} + \mbox{h.c.}
= \frac{C_{e\gamma}^{pr}}{\Lambda^2}
 (\bar l_p \sigma^{\mu\nu} e_r) \vp F_{\mu\nu} + \mbox{h.c.},
\end{align}
where $F_{\mu\nu}$ is the electromagnetic field-strength tensor, is
then the only term in the Dim-6 Lagrangian that induces a $l_i\to
l_j\gamma$ transition at tree level.

In the Feynman gauge, the combination of $Q_{e\vp}$ with the Dim-4 SM Yukawa terms gives
\begin{align}\label{eq:YukPlusEff}
&\mathcal{L}_{\rm Yukawa}+\mathcal{L}_{e\varphi}=\nonumber \\
=&\frac{v}{\sqrt{2}}\left(
-y_{pr}+\frac{v^2}{2\Lambda^2}C_{e\varphi}^{pr}
\right)\bar{e}_pe_r\nonumber \\
+&\frac{1}{\sqrt{2}}\left(
-y_{pr}+\frac{v^2}{2\Lambda^2}C_{e\varphi}^{pr}
\right)\bar{e}_pe_rh+
\frac{v^2}{\sqrt{2}\Lambda^2}C_{e\varphi}^{pr}\bar{e}_pe_rh\nonumber \\
+&\frac{i}{\sqrt{2}}\left(
-y_{pr}+\frac{v^2}{2\Lambda^2}C_{e\varphi}^{pr}
\right)\bar{e}_pe_r\widehat{Z}\nonumber \\
+&i\left(
-y_{pr}+\frac{v^2}{2\Lambda^2}C_{e\varphi}^{pr}
\right)\bar{e}_p\nu_r\widehat{W}^++\left[\dots\right].
\end{align}
From \Eqn{eq:YukPlusEff}, it is understood that any 3-point
off-diagonal interaction involving Goldstone bosons is not physical,
i.e. it can be removed by an orthogonal transformation\footnote{In
  this proceeding, any impact of such diagonalisation on the charged
  lepton eigenstates and their masses will be neglected.}. However,
this procedure leaves a residual term with a physical Higgs supporting
LFV currents, which is the only tree-level contribution to the $H\to
l_il_j$ transition.

Concerning the decay $Z\to l_i l_j$, such transition is triggered at
the tree level by the following operators: $Q_{eZ}$ (the dipole
contribution), $Q_{\vp l}^{(1)}$, $Q_{\vp l}^{(3)}$ and $Q_{\vp
  e}$. In addition, $\mu$ and $\tau$ three body decays are also
produced by the point-like four-fermion operators $Q_{ll}$, $Q_{le}$
and $Q_{ee}$.

The impact of $Q_{lequ}^{(1)}$ and $Q_{lequ}^{(3)}$ in the $l_i\to
l_j\gamma$ transition is of a different kind: they contribute in the
running of the dipole operators, hence they must be carefully included
when different energy scales are considered, as was shown in~\cite{Jenkins:2013zja,Jenkins:2013wua,Alonso:2013hga,Pruna:2014asa}.

\section{Branching Ratios}
\label{sec-3}
In the limit where $m_{1}\gg m_{2}$ and \emph{no correlation is
  considered among different operators}, the partial widths of the
aforementioned processes are the following:
\begin{itemize}
\item for the two-body decay $l_1^\pm \to l_2^\pm\gamma$, one has
\begin{align}\label{eq:meg}
&\Gamma(l_1^\pm \to l_2^\pm\gamma)=
\frac{m_1^3}{4\pi \Lambda^4}\left(|C_{TL}|^2+|C_{TR}|^2\right),
\end{align}
where the contributions to $C_{TL}$ and $C_{TR}$ (at the energy scale
$\lambda=m_Z$, in the assumption that the coefficients are real) are
given in Table~\ref{tab:res}, according to~\cite{Pruna:2014asa} (see also~\cite{Crivellin:2013hpa}). The coefficients $C_k^{(d)}$ in
Table~\ref{tab:res} are to be interpreted as renormalised in the
$\overline{\rm MS}$-scheme;
\item for the three-body decay $l_1^\pm \to l_2^\pm l_2^\mp l_2^\pm$, one has
\begin{align}
&\Gamma(l_1^+ \to l_2^+ l_2^- l_2^+)=
\Bigg(40 e^2 v^2 \left(\left|C_{e\gamma}^{12}\right|^2+\left|C_{e\gamma}^{21}\right|^2\right) \left(8 \ln\left[\frac{m_1}{m_2}\right]-11\right)\nonumber\\
&+\frac{2 m_1^4}{m_Z^2}\bigg(\left(5-20 s_W^2+36 s_W^4\right)
\left|C_{eZ}^{12}\right|^2\nonumber\\
&\qquad\quad + 4 \left(1-4 s_W^2+9 s_W^4\right) \left|C_{eZ}^{21}\right|^2\bigg)\nonumber \\
&+\frac{15 m_2^2 m_1^2 v^2 \left(
\left|C_{e\vp}^{12}\right|^2+\left|C_{e\vp}^{21}\right|^2\right)}{8 m_H^4}\nonumber \\
&+
10 m_1^2 \left(1-4 s_W^2+12 s_W^4\right) \left|C_{\vp e}^{12}\right|^2\nonumber \\
&+20 m_1^2 \left(1-4 s_W^2+6 s_W^4\right) \left(\left|C_{\vp l(1)}^{12}\right|^2+
\left|C_{\vp l(3)}^{12}\right|^2\right)+ \nonumber \\
&+10 m_1^2 \left(\left|C_{le}^{1112}\right|^2+\left|C_{le}^{1211}\right|^2\right)
\nonumber \\
&+80 m_1^2 \left(\left|C_{ee}^{1112}\right|^2+\left|C_{ll}^{1112}\right|^2\right)
\Bigg)\frac{m_1^3}{30 (8\pi)^3 \Lambda^4}, \label{l3l}
\end{align}
where the integration over the phase space of the photonic dipole
contribution gives rise to a logarithmic term, in agreement with~\cite{Kuno:1999jp}. The result in Eq.~(\ref{l3l})
is also valid for $\Gamma(l_1^- \to l_2^- l_2^+ l_2^-)$;
\item flavour-violating $Z$ decays can be parametrised at the tree
  level by means of the following four operators: 
\begin{align}\label{eq:Zem}
&\Gamma(Z\to l_1^\pm l_2^\mp)=\frac{m_Z^3 v^2}{12 \pi  \Lambda^4}
  \left(
  \left|C_{eZ}^{12}\right|^2+\left|C_{eZ}^{21}\right|^2\right.\nonumber\\ 
&\left.+   \left|C_{\vp e}^{12}\right|^2+\left|C_{\vp
    l(1)}^{12}\right|^2+\left|C_{\vp l(3)}^{12}\right|^2\right), 
\end{align}
and as one can see from the last equation, all of their contributions
occur at the same order. In Eq.~(\ref{eq:Zem}) we have summed over the
two possible final states, $l_1^+ l_2^-$ and $l_1^- l_2^+$;
\item for the Higgs boson decay $H\to l_1^\pm l_2^\mp$, one has
\begin{align}\label{eq:Hem}
\Gamma(H\to l_1^\pm l_2^\mp)=\frac{m_H
  v^4}{16\pi\Lambda^4}\left(\left|C_{e\vp}^{12}\right|^2+\left|C_{e\vp}^{21}\right|^2\right), 
\end{align}
where only one operator contributes at tree level. As for the $Z$
decays, in Eq.~(\ref{eq:Hem}) we have summed over the two possible
decays  $l_1^+ l_2^-$ and $l_1^- l_2^+$. 
\end{itemize}
\begin{table}[!ht] 
\centering
\renewcommand{\arraystretch}{1.2}
\btb{||c|c||} 
\hline 
\hline 
Op. &  $C_{TL}$ or 
$C_{TR} (1 \longleftrightarrow 2)$ \\
\hline 
\  & \  \\[-2.5ex]
$Q_{e\gamma}$ & 
$
\begin{aligned}
-C_{e\gamma}^{12}\frac{v}{\sqrt{2}}
\end{aligned}
$
 \\[2.5ex]
$Q_{eZ}$ & 
 $
\begin{aligned}
-C_{eZ}^{12}\frac{ e m_Z }{16 \sqrt{2} \pi ^2}
\left(3-6 c_W^2+4 c_W^2 \log\left[c_W^2\right]
\right)
\end{aligned}
$
 \\[2.5ex]
$Q_{\vp l}^{(1)}$ & 
 $
\begin{aligned}
-C_{\vp l(1)}^{12}\frac{e m_1 \left(1+s_W^2\right) }{24 \pi ^2}
\end{aligned}
$
 \\[2.5ex]
$Q_{\vp l}^{(3)}$ &
 $
\begin{aligned}
C_{\vp l(3)}^{12}\frac{e m_1 \left(3-2 s_W^2\right)}{48 \pi ^2}
\end{aligned}
$
 \\[2.5ex]
$Q_{\vp e}$ & 
$
\begin{aligned}
C_{\vp e}^{12}\frac{e m_2 \left(3-2 s_W^2\right) }{48 \pi ^2}
\end{aligned}
$
 \\[2.5ex]
$Q_{e\vp}$ & 
$
\begin{aligned}
C_{e\vp}^{12}\frac{e v\ {\rm Max}(m_1^2,m_2^2)}{96 \sqrt{2} m_H^2 \pi ^2} 
\left(4 +3 
\log\left[\frac{{\rm Max}(m_1^2,m_2^2)}{m_H^2}\right]
\right)
\end{aligned}
$
 \\[2.5ex]
$Q_{le}$ &  $
\begin{aligned}
\frac{e}{16 \pi ^2}\sum_{i=e,\mu,\tau}m_i C_{le}^{1ii2}
\end{aligned}
$ 
 \\[1ex]
 \hline
\hline
\etb
\caption{Leading order results (up to one loop) for the contributions
  of the various (real) Dim-6 operators to the $l_1\to l_2\gamma$ decay
  at the energy scale $\lambda=m_Z$. \label{tab:res}}
\end{table}

In order to numerically evaluate the BRs, the Particle Data Group
(PDG) values~\cite{Agashe:2014kda} for the physical constants and for
the $\mu$ and $\tau$ total decay width were used. 
%
\section{Results}
\label{sec-4}
Here, the elements collected in Section~\ref{sec-3} are combined with
the experimental limits on LFV transitions at high and low
energies. Since the category of transitions $l_i\to l_j\gamma$ was
studied at the one-loop level, the limits on $\mu\to e \gamma$ and
$\tau \to \mu\gamma$ from MEG and BaBar were interpreted at the energy
scale $\lambda=m_Z$ by means of the QED running of the Wilson
coefficients in Eq.~(\ref{eq:meg}). Being absolutely rigorous, also
the limits on three-body decays should be studied at the one-loop
level and reinterpreted at the $Z$-boson mass scale. However, it was
proved by direct computation that such running, in absence of
correlations among operators, only affects the limits at the $10\%$
level. Hence it is not relevant in the following discussion where
differences of orders of magnitude are involved.

\subsection{Limits at a fixed energy $\lambda=m_Z$}
\label{sec-4.1}
Under the assumption that only one operator at a time is non-vanishing
and the corresponding coefficients are real, the numerical limits of
Tables~\ref{tab:nresm}-\ref{tab:nrest} are obtained. Note that $C_{\vp
  e}^{21} = (C_{\vp e}^{12})^\ast \doteq C_{\vp e}^{12}$ (and similar
for $C_{\vp l(1)}$ and $C_{\vp l(3)}$) since we treat the coefficients
as real. We also have $C_{ee}^{prst} = C_{ee}^{stpr} =
(C_{ee}^{tsrp})^\ast \doteq C_{ee}^{tsrp}$ (and similar for $C_{ll}$),
where again the last step assumes the coefficients to be real.

\begin{table}[!ht] 
\centering
\renewcommand{\arraystretch}{1.2}
\begin{tabular}{||c||c|c|c||} 
\hline 
\hline 
Coeff. & $\mu^+\to e^+\gamma$  & $Z\to e^\pm\mu^\mp$ & $\mu^+\to e^+e^-e^+$  \\[-1ex]
{\scriptsize $\lambda=m_Z$} & {\scriptsize BR$\leq 5.7\cdot 10^{-13}$}
& 
{\scriptsize BR$\leq 7.5\cdot 10^{-7}$} & {\scriptsize BR$\leq 1.0\cdot 10^{-12}$}  \\
\hline 
$C_{e\gamma}^{21/12}$ &
\fcolorbox{red}{green}{$2.5\cdot 10^{-16}$} & 
\ &
$3.8\cdot 10^{-15}$  \\
$C_{eZ}^{21/12}$ &
\fcolorbox{red}{green}{$1.4\cdot10^{-13}$} & 
$3.9\cdot 10^{-8}$ & 
$4.0\cdot 10^{-8}$ \\
$C_{\varphi l(1)}^{12}$ &
$2.6\cdot10^{-10}$ & 
$3.9\cdot10^{-8}$ & 
\fcolorbox{red}{green}{$3.5\cdot10^{-11}$} \\
$C_{\varphi l(3)}^{12}$ &
$2.5\cdot10^{-10}$ &
$3.9\cdot10^{-8}$ & 
\fcolorbox{red}{green}{$3.5\cdot10^{-11}$} \\
$C_{\varphi e}^{12}$ & 
$2.5\cdot10^{-10}$ &
$3.9\cdot10^{-8}$ & 
\fcolorbox{red}{green}{$3.7\cdot10^{-11}$} \\
$C_{e\varphi}^{21/12}$ &
\fcolorbox{red}{green}{$2.8\cdot10^{-8}$}&  &
$8.7\cdot10^{-6}$\\
$C_{le}^{2111/1112}$ &
$4.4\cdot 10^{-8}$&  &
\fcolorbox{red}{green}{$3.1\cdot 10^{-11}$} \\
$C_{le}^{2221/1222}$ &\
\fcolorbox{red}{green}{$2.1\cdot 10^{-10}$} &  & \\
$C_{le}^{2331/1332}$ & 
\fcolorbox{red}{green}{$1.2\cdot 10^{-11}$} &  & \\
$C_{ee}^{2111}$ & &  &
\fcolorbox{red}{green}{$1.1\cdot 10^{-11}$} \\
$C_{ll}^{2111}$ & &  &
\fcolorbox{red}{green}{$1.1\cdot 10^{-11}$} \\[1ex]
 \hline 
\hline
\end{tabular}
\caption{Limits on the Wilson coefficients in [GeV]$^{-2}$
  contributing to the LFV muonic transitions. $\mu\to 3e$ will lead to better limits on $C_{eZ}$ once NLO corrections are taken into account. \label{tab:nresm}}
\end{table}

Focusing on LFV muonic transitions, the first thing to notice in
Table~\ref{tab:nresm} is that the best limits are currently always
obtained from the constraints established by low energy
experiments. The constraints from LFV decays $Z\to e^\pm \mu^\mp$
on the coefficients of the operators $Q_{eZ}$, $Q_{\vp l(1,3)}$ and
$Q_{\vp e}$ are considerably less stringent than from $\mu^+\to
e^+\gamma$ and $\mu^+\to e^+e^-e^+$. 

Even if at a future high energy lepton colliders (e.g. FCC-ee~\cite{Gomez-Ceballos:2013zzn}) the LEP limits for BR$(Z\to e^\pm
\mu^\mp)$ could be improved by several orders of magnitude it is still
the case that the information one can get from such future machine on
$Z\to e^\pm \mu^\mp$ is only complementary, and surely never
competitive with the planned future low energy experiments. Indeed, if
no correlation is assumed, from Eq.~(\ref{eq:Zem}) one deduces that in
order to reach a probing power equivalent to the current low energy
experiments, the required branching ratios for $Z\to e^\pm \mu^\mp$
are BR$\leq 9.5\cdot 10^{-18}$ for $C_{eZ}^{21}$, BR$\leq 5.9\cdot
10^{-13}$ for $C_{\vp l(1,3)}^{21}$ and BR$\leq 6.6\cdot 10^{-13}$
for $C_{\vp e}^{21}$, respectively. Hence, the operator
$Q_{eZ}^{21}$ will be out of the reach of FCC-ee, while the
coefficients of $Q_{\vp l(1)}^{12}$, $Q_{\vp l(3)}^{12}$ and
$Q_{\vp e}^{12}$ could be constrained with (more or less) the same
precision of the dated SINDRUM experiment. The interplay between a
high-luminosity $Z$-factory and low-energy experiments has also been
studied with concrete BSM models (see e.g.~\cite{Abada:2014cca}). 

Comparing the limits obtained from $\mu^+\to e^+\gamma$ with $\mu^+\to
e^+e^-e^+$ it is clear that both experiments provide very valuable
information. The standard statement is that $\mu\to 3 e$ is more
powerful in constraining four-fermion operators (contact interactions)
whereas $\mu\to e\gamma$ is more sensitive to the dipole
operators. However, a closer look reveals that this is only partially
true. Some four-fermion operators have stringent limits due to $\mu\to
e\gamma$ (at one loop) but are not constrained by $\mu\to 3 e$ (at
tree level). Furthermore, in the long term $\mu\to 3 e$ might well be
competitive in putting limits on the photon dipole operator. The BR
for $\mu\to 3 e$ is expected to be improved by several orders of
magnitude. A limit BR$(\mu\to 3 e) \leq 4\cdot 10^{-15}$ is as
constraining to $C_{e\gamma}^{21}$ as the current BR$(\mu\to e\gamma)
\leq 5.7\cdot 10^{-13}$. Of course, the BR for $\mu\to e \gamma$ is
also expected to improve and in the longer term, there will also be
very strong constraints from future muon conversion
experiments. Regarding the operators with a Higgs, $\mu\to 3 e$
provides the stronger limits except for $C_{e\vp}$. However, even
stronger limits on this coefficient can be obtained by looking at muon
conversion~\cite{Petrov:2013vka,Crivellin:2014cta}.  Let us mention again that our
limit on $C_{e\vp}$ is indicative at best, since we have not taken
into account numerically important two-loop contributions from
Barr-Zee diagrams~\cite{Barr:1990vd, Chang:1993kw}. We will come back
to this when discussing tau decays.

Note that the limits for $C_{ll}^{2111}$ and $C_{ee}^{2111}$
differ by a factor 2 from the corresponding limits given
in~\cite{Feruglio:2015yua}, since we write the terms containing
$Q_{ll}$ (and $Q_{ee}$) in the Lagrangian as
\begin{align} 
\mathcal{L}_{ll} = &\sum_{prst} \frac{C_{ll}^{prst}}{\Lambda^2}
(\bar l_p \gamma_\nu l_r)(\bar l_s \gamma^\nu l_t) 
\nonumber \\
&= 2 \frac{C_{ll}^{2111} }{\Lambda^2}
(\bar l_2 \gamma_\nu l_1)(\bar l_1 \gamma^\nu l_1)
+ \left[\dots\right],
\end{align}
where we exploited the symmetry of the operator. 
\begin{table}[!ht] 
\centering
\renewcommand{\arraystretch}{1.2}
\begin{tabular}{||c||c|c|c||} 
\hline 
\hline 
Coeff. & $\tau^+\to \mu^+\gamma$  & $Z\to \mu^\pm\tau^\mp$ & $\tau^+\to \mu^+\mu^-\mu^+$  \\[-1ex]
{\scriptsize $\lambda=m_Z$} & {\scriptsize BR$\leq 4.4\cdot 10^{-8}$}  & {\scriptsize BR$\leq 1.2\cdot 10^{-5}$} & {\scriptsize BR$\leq 2.1\cdot 10^{-8}$}  \\
\hline 
$C_{e\gamma}^{32/23}$ &
\fcolorbox{red}{green}{$2.7\cdot 10^{-12}$} & 
\ &
$3.8\cdot 10^{-11}$  \\
$C_{eZ}^{32/23}$ &
\fcolorbox{red}{green}{$1.5\cdot10^{-9}$} & 
$1.5\cdot 10^{-7}$ & 
$8.7\cdot 10^{-7}$ \\
$C_{\varphi l(1)}^{23}$ &
$1.7\cdot10^{-7}$ & 
$1.5\cdot10^{-7}$ & 
\fcolorbox{red}{green}{$1.3\cdot10^{-8}$} \\
$C_{\varphi l(3)}^{23}$ &
$1.6\cdot10^{-7}$ &
$1.5\cdot10^{-7}$ & 
\fcolorbox{red}{green}{$1.3\cdot10^{-8}$} \\
$C_{\varphi e}^{23}$ & 
$1.6\cdot10^{-7}$ &
$1.5\cdot10^{-7}$ & 
\fcolorbox{red}{green}{$1.3\cdot10^{-8}$}\\
\cline{1-1}
\cline{3-3}
Coeff. &   & $H\to \mu^\pm\tau^\mp$ &   \\[-1ex]
{\scriptsize $\lambda=m_Z$} &   & {\scriptsize BR$\leq 1.8\cdot 10^{-2}$} &   \\
\cline{1-1}
\cline{3-3}
$C_{e\varphi}^{32/23}$ &
$1.9\cdot10^{-6}$& \fcolorbox{red}{green}{$9.0\cdot10^{-8}$} &
$1.6\cdot10^{-5}$\\
$C_{le}^{3112/2113}$ &
\fcolorbox{red}{green}{$4.8\cdot 10^{-4}$}&  &
 \\
$C_{le}^{3222/2223}$ &\
$2.3\cdot 10^{-6}$ &  &
\fcolorbox{red}{green}{$1.1\cdot 10^{-8}$} \\
$C_{le}^{3222/2223}$ & 
\fcolorbox{red}{green}{$1.4\cdot 10^{-7}$} &  & \\
$C_{ee}^{3222}$ & &  &
\fcolorbox{red}{green}{$4.0\cdot 10^{-9}$} \\
$C_{ll}^{3222}$ & &  &
\fcolorbox{red}{green}{$4.0\cdot 10^{-9}$} \\[1ex]
 \hline 
\hline
\end{tabular}
\caption{Limits on the Wilson coefficients in [GeV]$^{-2}$
  contributing to the LFV tauonic transitions. \label{tab:nrest}}
\end{table}

The analysis can easily be adapted to the tau sector and the
corresponding limits on the coefficients are given in
Table~\ref{tab:nrest}. We should note that similar results can be
found in the literature~\cite{Dassinger:2007ru, Crivellin:2013hpa,Celis:2014roa,
  Celis:2014asa} and a review of studies of the
tauonic limits can be found e.g. in~\cite{Feruglio:2015yua}. The
pattern is very similar to muonic decays. However, the limits from LFV
$Z$ decays play a more important role and if they were to be improved
by several orders of magnitude at a future high energy lepton
colliders, they could provide the most stringent limits on $C_{\vp
  l(1)}^{23}$, $C_{\vp l(3)}^{23}$ and $C_{\vp e}^{23}$.

A new feature of the tau sector is the appearance of the $H\to\tau\mu$
decay, which has caused a huge theoretical
activity~\cite{Blankenburg:2012ex, Harnik:2012pb, Kopp:2014rva,Sierra:2014nqa,
  deLima:2015pqa, Dorsner:2015mja, Omura:2015nja}. At tree level, the
only operator that induces such a LFV decay in the Lagrangian
Eq.~(\ref{eq:bw}) is $Q_{e\varphi}^{32/23}$. The non-observation of
this decay at the LHC is placing a limit on the $C_{e\varphi}^{32/23}$
coefficient that is more stringent then any limit extracted from the
low energy experiment. This decay has been considered extensively in
the literature, often with phenomenologically motivated
Lagrangians~\cite{Blankenburg:2012ex}, where the LFV interactions are
given in terms of generalised Yukawa matrices. In terms of Dim-6
operators, the deviations from the SM Yukawa interactions are induced
by the operator $Q_{e\varphi}$, as can be seen in
Eq.~(\ref{eq:YukPlusEff}). Our limits on $C_{e\varphi}^{32/23}$ in
Table~\ref{tab:nrest} indicate that $H\to\tau\mu$ provides the
strongest constraint, followed by $\tau\to\mu\gamma$ and $\tau\to
3\mu$. This is in qualitative agreement with~\cite{Harnik:2012pb}. However, we should mention once more that the
limits extracted here from low-energy observables do not consider the
impact of Barr-Zee type two-loop contributions, which is
particularly important in this case~\cite{ Harnik:2012pb}.

In Table~\ref{tab:nrest}, the ATLAS \& CMS data on $H\to \tau\mu$ are
considered only in their function of placing new limits on such a
transition. However, if one wants to consider the possibility that the
picture corresponds to a signal compatible with $\rm{BR}(H\to
\tau\mu)=0.84\%$ (with a significance of $2.4$ $\sigma$), then a
trivial computation leads to $C_{e\varphi}^{32/23}/\Lambda^2\sim
8.7\cdot 10^{-8}$~GeV$^{-2}$ or, in the same way, to $\Lambda\sim 3.4
\sqrt{C_{e\varphi}^{32/23}}$~TeV. From this, it follows that an
effective coefficient that preserves the perturbative behaviour
$C_{e\varphi}^{32/23}\le 1$ would imply a limit on the UV
completion scale $\Lambda\ge 3.4$ TeV, that in turn preserves the
perturbative behaviour of the expansion in Eq.~(\ref{eq:bw}), being
$C_{e\varphi}(v/\Lambda)^2\le \mathcal{O}(10^{-2})$. Here, no further
speculation will be carried out about the possible nature of the UV
completion and we refer to e.g.~\cite{Altmannshofer:2015esa,
  Crivellin:2015lwa} for a deeper discussion on the topic.

\subsection{Limits from RGE}
\label{sec-4.2}
In the previous subsection we have given limits that were obtained by
comparing NLO (or LO) calculations using Eq.~(\ref{eq:bw}) to
experimental limits on branching ratios. However, the most direct link
of the Wilson coefficients $C_k^{(d)}$ of Eq.~(\ref{eq:bw}) to the
(unknown) UV complete theory is at the large scale $\Lambda$. In order
to obtain the Wilson coefficients at the low scale ($m_Z$ or $m_l$)
that are relevant for the branching ratios, RG evolution has to be
applied. This has two effects. First, the limits on $C_k^{(d)}(m_Z)$
will be translated into limits $C_k^{(d)}(\Lambda)$. Second, and more
important, qualitatively new effects happen. It might well be
possible, that a certain UV complete theory does result in a
vanishing coefficient $C_{e\gamma}(\Lambda) = 0$ at the large
scale. Nevertheless, the corresponding theory can lead to a LFV decay
$l^i\to l^j\gamma$. In fact, other operators that are generated by the
UV-complete theory at $\Lambda$ can mix under RG evolution with
$Q_{e\gamma}$ and result in a non-vanishing Wilson coefficient
$C_{e\gamma}(m_z) \neq 0$. 

In principle, the complete set of Dim-6 operators has to be taken into
account to study $C_{e\gamma}(m_z)$. Restricting ourselves to
operators that either directly or indirectly mix into $Q_{e\gamma}$, the RG evolution of the corresponding coefficients is given in~\cite{Pruna:2014asa}:
\begin{align} 
16\pi^2\frac{\partial C_{e\gamma}^{i j}}{\partial\log{\lambda}}
&\simeq \left(\frac{47 e^2}{3}+\frac{e^2}{4 c_W^2}
-\frac{9 e^2}{4 s_W^2}+3 Y_t^2\right)C_{e\gamma}^{i j}\nonumber \\
&+6e^2\left(\frac{c_W }{s_W}-\frac{s_W}{c_W}\right)C_{eZ}^{i j}
+16 e Y_tC^{(3)}_{i j 33},\nonumber \\
16\pi^2\frac{\partial C_{eZ}^{i j}}{\partial \log{\lambda}}
&\simeq -\frac{2e^2}{3}\left(\frac{2 c_W }{s_W}
+\frac{31 s_W}{c_W}\right)C_{e\gamma}^{i j}\nonumber \\
&+2e\left(\frac{3 c_W}{s_W}-\frac{5 s_W}{c_W}\right)Y_tC^{(3)}_{i j 33}\nonumber \\
&+\left(-\frac{47 e^2}{3}+\frac{151 e^2}{12 c_W^2}
-\frac{11 e^2}{12 s_W^2}+3 Y_t^2\right)C_{eZ}^{i j},\nonumber\\
16\pi^2\frac{\partial C^{(3)}_{i j 33}}{\partial\log{\lambda}}&\simeq 
\frac{7 e Y_t}{3}C_{e\gamma}^{i j}+
\frac{eY_t}{2}\left(
\frac{3 c_W }{ s_W}-\frac{5 s_W }{3 c_W}
\right)C_{eZ}^{i j}+\nonumber \\
&+\left(
\frac{2 e^2}{9 c_W^2}-\frac{3 e^2}{s_W^2}+\frac{3 Y_t^2}{2}
+\frac{8 g_S^2}{3}\right)C^{(3)}_{i j33}\nonumber \\
&+
\frac{e^2}{8}  \left(\frac{5}{c_W^2}
+\frac{3}{s_W^2}\right)C^{(1)}_{i j33},\nonumber \\
16\pi^2\frac{\partial C^{(1)}_{i j33}}{\partial\log{\lambda}}&\simeq
\left(\frac{30 e^2}{c_W^2}+\frac{18 e^2}{s_W^2}\right)C^{(3)}_{i j33}\nonumber \\
&+
\left(
-\frac{11 e^2}{3 c_W^2}+\frac{15 Y_t^2}{2}-8 g_S^2
\right)C^{(1)}_{i j33}. \label{eq:rge}
\end{align}
We would like to point out the role of $C^{(3)}_{i j 33}$ and
$C^{(1)}_{i j33}$, the Wilson coefficients of the operators
$Q_{lequ}^{(3)}$ and $Q_{lequ}^{(1)}$ respectively. These operators
enter in the running of $C_{e\gamma}^{i j}$ with a coefficient
proportional to the Yukawa coupling. Hence we restrict ourselves to the
contribution of the top quarks in Eq.~(\ref{eq:rge}). Note that
$Q_{lequ}^{(3)}$ mixes directly into $Q_{e\gamma}$, resulting in a
rather stringent limit on $C^{(3)}_{i j 33}(\Lambda)$ from ${\rm
  BR}(l_i\to l_j\gamma)$. The constraint on $C^{(1)}_{ij33}(\Lambda)$
is weaker, since $Q_{lequ}^{(1)}$ enters only indirectly, i.e. through
$Q_{lequ}^{(3)}$.

In Tables~\ref{tab:rge_res} and~\ref{tab:rge_tmres} we list the limits
on the various coefficients of Eq.~(\ref{eq:rge}) obtained from
$\mu\to e\gamma$ and $\tau\to \mu\gamma$, respectively. Obviously,
very similar limits can be derived for $\tau\to e\gamma$. The limits
are given for several choices of the large scale $\Lambda$. As can be
seen from the entries of Tables~\ref{tab:rge_res} and~\ref{tab:rge_tmres}, the effect of the RG evolution is very modest for
$C_{e\gamma}^{ij}$ and more pronounced for $C_{eZ}^{ij}$. We also
point out that the limits obtained for $C_{eZ}^{ij}$ through the
effect of the RG evolution is considerably more stringent than the
corresponding limit from a standard NLO calculation, as can be seen by
comparing the entries in Tables~\ref{tab:nresm} and~\ref{tab:nrest}
with those in Tables~\ref{tab:rge_res} and~\ref{tab:rge_tmres},
respectively.  More important, however, is the fact that without
taking the RG into account, it would be impossible to get meaningful
limits on the coefficients of $C^{(3)}_{ij33}(\Lambda)$ and
$C^{(1)}_{ij33}(\Lambda)$. For illustration, we have also included in
the Tables limits on these coefficients in the charm sector,
$C^{(3)}_{ij22}(\Lambda)$ and $C^{(1)}_{ij22}(\Lambda)$. Since the
corresponding Yukawa coupling is smaller, the limits are less
constraining.
\begin{table}[t] 
\centering
\renewcommand{\arraystretch}{1.2}
\btb{||c||c|c|c||} 
\hline 
\hline
\multicolumn{4}{||c||}{$\boldsymbol{\mu\to e\gamma}$}\\ 
\hline 
Coeff. & $\Lambda = 10^3$ GeV &
$\Lambda = 10^5$ GeV & $\Lambda = 10^7$ GeV \\
\hline 
$C_{e\gamma}^{21}$ &  $2.7\cdot 10^{-10}$ & $2.9\cdot 10^{-6}$ & $3.1\cdot 10^{-2}$ \\
$C_{eZ}^{21}$ & $2.5\cdot 10^{-8}$ & $1.0\cdot 10^{-4}$ &  $7.1\cdot 10^{-1}$  \\
$C^{2133}_{lequ(3)}$ & $3.6\cdot 10^{-9}$ & $1.4\cdot 10^{-5}$ & $9.8\cdot 10^{-2}$ \\
$C^{2133}_{lequ(1)}$ & $1.9\cdot 10^{-6}$ & $2.5\cdot 10^{-3}$ & n/a  \\
$C^{2122}_{lequ(3)}$ & $4.8\cdot 10^{-7}$ & $1.9\cdot 10^{-3}$ & n/a \\
$C^{2122}_{lequ(1)}$ & $2.6\cdot 10^{-4}$ & $3.3\cdot 10^{-1}$ & n/a \\
 \hline 
\hline
\etb
\caption{Limits on the Wilson coefficients defined at the scale
  $\lambda=\Lambda$ for three choices of $\Lambda =10^3$, $10^5$,
  $10^7$ GeV. \label{tab:rge_res} }
\end{table}
\begin{table}[!ht] 
\centering
\renewcommand{\arraystretch}{1.2}
\btb{||c||c|c|c||} 
\hline 
\hline
\multicolumn{4}{||c||}{$\boldsymbol{\tau\to \mu\gamma}$}\\ 
\hline 
Coeff. & $\Lambda = 10^3$ GeV &
$\Lambda = 10^4$ GeV & $\Lambda = 10^5$ GeV \\
\hline 
$C_{e\gamma}^{32}$ & $3.0\cdot10^{-6}$ & $3.1\cdot10^{-4}$ & $3.2\cdot10^{-2}$ \\
$C_{eZ}^{32}$ &  $2.8\cdot10^{-4}$   & $1.5\cdot10^{-2}$ & $\sim 1.1$ \\
$C^{3233}_{lequ(3)}$ & $4.0\cdot10^{-5}$ & $2.2\cdot10^{-3}$  & $1.6\cdot10^{-1}$  \\
$C^{3233}_{lequ(1)}$ &  $2.1\cdot10^{-2}$ &  $5.9\cdot10^{-1}$ &  n/a \\
$C^{3222}_{lequ(3)}$ &  $5.4\cdot10^{-3}$ &  $3.0\cdot10^{-1}$ & n/a  \\
$C^{3222}_{lequ(1)}$ & $\sim 2.8$  &  n/a  &  n/a \\[1ex]
 \hline 
\hline
\etb
\caption{Limits on the Wilson coefficients defined at the scale
  $\lambda=\Lambda$ for three choices of $\Lambda =10^3$, $10^4$,
  $10^5$ GeV. \label{tab:rge_tmres}}
\end{table}
\section{Conclusion}
\label{sec-5}

In these proceedings we have described the first steps towards taking
the Standard Model effective field theory, that is the Lagrangian
given in Eq.~(\ref{eq:bw}), seriously as a quantum field theory. This
allows us to perform well defined calculations beyond leading order in
the Standard Model couplings and apply renormalisation group
techniques. At the current stage, the motivation to go beyond leading
order is not to obtain limits on the coefficients that are more
precise by $\sim$10\%. Rather, performing NLO calculations and taking
RG evolution effects into account, we obtain qualitatively new
results. This is the case in particular when considering the very
stringent limits of low-energy LFV experiments.

We take the attitude that $\Lambda$, the scale of new physics
responsible for LFV effects is much larger than the electroweak
scale. Hence we restrict ourselves to terms suppressed by at most two
powers of $\Lambda$, i.e. to single insertions of Dim-6 operators.
Ideally, the coefficients of these operators could all be measured
experimentally. This would provide a huge amount of information about
the nature of the UV complete theory valid beyond the scale $\Lambda$.
In practise, a relatively small number of experiments related to LFV
in the charged sector will be carried out at energy scales well below
$\Lambda$ and typically result in exclusion limits.  In order to
extract the maximal possible amount of information from these
experiments it is essential to perform computations beyond tree level
and take RG evolution into account. 

A complete experimental determination of the various coefficients is
certainly not a realistic prospect, but even partial information on
the coefficients will provide extremely useful guidance for the search
of the UV complete theory. However, for this it is essential to know
$C_k^{(6)}(\Lambda)$, the coefficients at the high scale $\Lambda$. It
is at this scale where there is the most direct link between the
effective theory and the underlying theory.
Even if the underlying theory does not produce an operator $Q_{e\gamma}$ when integrating out the heavy modes it can still result in the LFV decay $\mu\to e\gamma$.
To describe this effect,
the inclusion of RG evolution is at least as important as the
inclusion of one-loop corrections. In particular, if $\Lambda$ is
considered to be far above the electroweak scale, RG evolution
leads to qualitatively new effects and can lead to substantial
improvements in the limits that are extracted from low-energy
measurements. 

In the results presented here we have considered the decay $l_i\to l_j \gamma$
at NLO and have taken RG effects into account. However, other
processes such as muon conversion have been completely ignored and for
others, e.g. the decay $l_i\to 3\, l_j$ only a tree-level analysis has
been made. Furthermore, the rather unrealistic assumption has been
made that only a single coefficient at a time is non-vanishing.
Clearly, there is considerable room for improvement. We are convinced
that with a more complete analysis, LFV processes can play an even
more important role in the search for physics beyond the Standard
Model. 

\begin{acknowledgement}
The work of GMP is supported by the Swiss National Science
Foundation (SNF) under contract 200021\_160156.
\end{acknowledgement}

\bibliographystyle{woc}
\bibliography{biblio}
%
%
%
%

\end{document}